\documentclass[twocolumn]{aastex631}

\usepackage{graphicx}	
\usepackage{amsmath}	
\usepackage{color}
\usepackage[normalem]{ulem}
\usepackage{units}
\usepackage{nicefrac}

\newcommand{\SPA}{School of Physics and Astronomy, Monash University, Clayton VIC 3800, Australia}
\newcommand{\OzGravMonash}{OzGrav: The ARC Centre of Excellence for Gravitational Wave Discovery, Clayton VIC 3800, Australia}

\begin{document}

\title[Are all models wrong?]{Are all models wrong?
Falsifying binary formation models in gravitational-wave astronomy using exceptional events}

\author{Lachlan Passenger}
\email{lachlan.passenger1@monash.edu}
\affiliation{\SPA}                                                                                                                               
\affiliation{\OzGravMonash}                                                           

\author{Eric Thrane}
\affiliation{\SPA}
\affiliation{\OzGravMonash}

\author{Paul Lasky}
\affiliation{\SPA}                                                                       \affiliation{\OzGravMonash}

\author{Ethan Payne}
\affiliation{Department of Physics, California Institute of Technology, Pasadena, CA 91125, USA}
\affiliation{LIGO Laboratory, California Institute of Technology, Pasadena, CA 91125, USA}                    

\author{Simon Stevenson}
\affiliation{\OzGravMonash}                                                        
\affiliation{Centre for Astrophysics and Supercomputing, Swinburne University of Technology, Hawthorn, VIC 3122, Australia}    

\author{Ben Farr}
\affiliation{Institute for Fundamental Science, Department of Physics, University of Oregon, Eugene, OR 97403, USA}





\begin{abstract}
As the catalogue of gravitational-wave transients grows, several entries appear ``exceptional'' within the population.
Tipping the scales with a total mass of $\sim 150 M_\odot$, GW190521 likely contained black holes in the pair-instability mass gap.
The event GW190814, meanwhile, is unusual for its extreme mass ratio and the mass of its secondary component.
A growing model-building industry has emerged to provide explanations for such exceptional events, and Bayesian model selection is frequently used to determine the most informative model.
However, Bayesian methods can only take us so far.
They provide no answer to the question: \textit{does our model provide an adequate explanation for exceptional events in the data?}
If none of the models we are testing provide an adequate explanation, then it is not enough to simply rank our existing models---we need new ones.
In this paper, we introduce a method to answer this question with a frequentist $p$-value.
We apply the method to different models that have been suggested to explain the unusually massive event GW190521: hierarchical mergers in active galactic nuclei and globular clusters.
We show that some (but not all) of these models provide adequate explanations for exceptionally massive events like GW190521.
\end{abstract}




\section{Introduction}
 \label{intro}
The LIGO \citep{theligoscientificcollaborationAdvancedLIGO2015}, Virgo \citep{acerneseAdvancedVirgo2nd2015a} and KAGRA \citep{akutsuKAGRAGenerationInterferometric2019} gravitational-wave collaborations have published $\sim100$ confident compact-binary mergers so far (\cite{ligoscientificcollaborationGWTC3CompactBinary2023}). 
Several of these events exhibit unusual properties. 
One such exceptional event is GW190521 \citep{ligoscientificcollaborationandvirgocollaborationGW190521BinaryBlack2020a}, with component masses  $m_1 = 85^{+21}_{-14}\text{ } M_\odot$ and $m_2 = 66^{+17}_{-18}\text{ } M_\odot$ ($90\%$ credible intervals). 
The formation of these high-mass black holes is difficult to explain in field-binary models due to pair-instability processes \citep[e.g.,][]{ hegerNucleosyntheticSignaturePopulation2002, woosleyPulsationalPairInstability2007, 
belczynskiEffectPairinstabilityMass2016b, woosleyPulsationalPairinstabilitySupernovae2017,woosleyEvolutionMassiveHelium2019, woosleyPairinstabilityMassGap2021}, which is expected to produce a black-hole mass gap of approximately $50-135 \text{ } M_{\odot}$. 
However, the precise bounds of this gap are unknown, due especially to uncertainties in nuclear-reaction rates \citep[e.g.,][]{farmerMindGapLocation2019, costaFormationGW190521Stellar2021} as well as envelope retention and mass fallback upon core collapse \citep[e.g.,][]{winchPredictingHeaviestBlack2024}.

Several competing formation channels have been proposed to produce extreme-mass events like GW190521: hierarchical mergers in dense star clusters \citep[e.g.,][]{fragioneOriginGW190521likeEvents2020, romero-shawGW190521OrbitalEccentricity2020c, mapelliHierarchicalBlackHole2021, dallamicoGW190521FormationThreebody2021a, arca-seddaBreachingLimitFormation2021, liuHierarchicalBlackHole2021, kimballEvidenceHierarchicalBlack2021} or in accretion disks around active galactic nuclei (AGN) \citep[e.g.,][]{tagawaMassgapMergersActive2021, palmeseLIGOVirgoBlack2021, samsingAGNPotentialFactories2022, vajpeyiMeasuringPropertiesActive2022, mortonGW190521BinaryBlack2023}, the binary evolution of Population~III stars \citep[e.g.,][]{liuPopulationIIIOrigin2020, safarzadehFormationGW190521Gas2020, kinugawaFormationBinaryBlack2021, tanikawaPopulationIIIBinary2021} or mergers of primordial black holes \citep[e.g.,][]{delucaGW190521MassGap2021, clesseGW190425GW190521GW1908142022, chenConfrontingPrimordialBlack2022}. 

There are different ways to determine whether a model is consistent with a catalogue of gravitational-wave events, each one corresponding to a different question.
These questions can be subtly different, emphasising different ways in which a model can fail to account for the data.
Thus, we advocate a multi-pronged approach, which considers a variety of complementary questions. 
For example, \citet{mouldOneManyComparing2023} calculate Bayes factors between different population models for the same event, and thus obtain a relative measure of a models suitability. 
Meanwhile, \citet[]{essickProbingExtremalGravitationalWave2022} propose an extension to leave-one-out outlier tests, which emphasises the self-consistency of parameterised models analysing different subsets of data.
Finally, \citet{Lstroke} introduce the concept of the ``maximum population likelihood'' as a tool for testing a model's overall goodness of fit for a gravitational-wave catalogue.

In this paper, we build on the foundation laid by the above studies, but explore a different aspect of model testing. We assess whether a particular model can plausibly explain the most exceptional events observed, without directly comparing it to other models.
In particular, we build on the work of \citet{fishbachMostMassiveBinary2020a}, which outlines how one may perform a posterior predictive check on the most massive event observed, to check consistency with an inferred population model.\footnote{We compare and contrast our method with the technique from \cite{fishbachDoesMatterMatter2020a} at the top of Section~\ref{discussion}.} We extend this method to account for the measurement uncertainty associated with high-mass events by factoring in the posterior distributions of extreme-event parameters, rather than using their maximum likelihood estimates.

We consider many simulated population catalogues, drawn from the model of interest. We take extremal events from these catalogues and calculate a Bayesian evidence. By computing the Bayesian evidence for many extremal events from different catalogue realizations, we construct a distribution to which we may compare. 
We use this distribution to calculate a statistic in the form of a $p$-value, which describes how consistent the observed exceptional event is with extremal simulated events.

In Section \ref{method}, we develop the statistical framework used in our analysis. In Section \ref{results}, we demonstrate a use of our method by assessing the capability of several AGN models and one globular cluster model with hierarchical mergers to produce an event as extreme in total mass as GW190521. We find that it is feasible for both AGN models with sufficiently high natal black hole masses and globular cluster models with zero natal black hole spins to produce GW190521-like events. This suggests that population models that allow for hierarchical mergers can explain the most massive gravitational-wave events we have observed thus far, and are capable of bridging the proposed pair-instability mass gap \citep[see also, e.g.,][]{fragioneOriginGW190521likeEvents2020, kimballEvidenceHierarchicalBlack2021, anagnostouRepeatedMergersBlack2022}.

\section{Method} \label{method}
Our starting point is some population model of interest $M$ that defines the prior distribution for parameter(s) $\theta$,
\begin{align}\label{eq:prior}
    \pi(\theta | M) ,
\end{align}
for example, the distribution for total mass of binary black hole systems.
This distribution, in turn, implies a distribution of detected events, which may have a different shape due to selection effects,
\begin{align}\label{eq:detected}
    \pi(\theta | M, \text{det}) \propto 
    \pi(\theta | M) \, p_\text{det}(\theta) .
\end{align}
Here, $p_\text{det}(\theta)$ is the probability that an event with parameters $\theta$ is detected.

Since we are interested in extremal events, we next consider a catalogue consisting of $N$ events.
Each catalogue has an apparently most extreme event such that the maximum likelihood estimator for that event $\widehat\theta$ is larger (or smaller) than all other events,
\begin{align}
    \widehat\theta_i \lessgtr 
    \widehat\theta_{j \neq i} .
\end{align}
There is some distribution of $\theta$ for these apparently most extreme events,
\begin{align}\label{eq:most_massive_detected}
    \pi( \theta_\text{ext} | M, \text{det}, N) .
\end{align}
For example, this distribution could represent the distribution of total mass for the apparently most-massive event in a catalogue of $N=100$ events.

In the limit where $\theta_\text{ext}$ is measured precisely, then the maximum likelihood estimator approaches the true parameter value
\begin{align}
    \widehat\theta_\text{ext} \rightarrow \theta_\text{ext} ,
\end{align}
and so the apparently most extreme event \textit{is} the most extreme event.\footnote{In this limit,
\begin{align}
    \pi( \theta | M, \text{det}, N) = & 
    \Phi(\theta | M, \text{det})^{N-1} \, 
    \pi(\theta | M, \text{det}) ,
\end{align}
where $\Phi$ is a cumulative density function
\begin{align}
    \Phi(\theta | M, \text{det}) = \int_{\theta_\text{min}}^\theta d\theta' \,
    \pi(\theta' | M, \text{det}) .
\end{align}
}
In this limit, one may calculate a $p$-value for an extremal event with $\widehat\theta_\text{ext}$ in a catalogue with $N$ events, as in \citet{fishbachMostMassiveBinary2020a},
\begin{align}\label{eq:naive-p}
    p \equiv &
    \int_{\Gamma_1} d\theta \, 
    \pi(\theta_\text{ext} | M, \text{det}, N) \\
    \Gamma_1 = & \bigg\{\pi(\theta_\text{ext}| M, \text{det}, N) < \pi(\widehat\theta_\text{ext} | M, \text{det}, N)\bigg\} .
\end{align}
This $p$-value corresponds to the fraction of draws from $\pi(\theta_\text{ext}|M,\text{det}, N)$ that produce a prior probability density smaller than what we obtain for the observed value of $\widehat\theta_\text{ext}$.
If the observed value of total mass is unusual---either too small or too big---then $\pi(\widehat\theta_\text{ext}|M,\text{det},N)$ will be small, and so the resulting $p$-value will be accordingly small.
On the other hand, if the observed total mass is typical given the model, then $\pi(\widehat\theta_\text{ext}|M,\text{det},N)$ is large, and so $p$ is ${\cal O}(1)$.

In reality, we do not measure $\theta_\text{ext}$ precisely.
Rather, each measurement is characterised with a precision determined by the likelihood function ${\cal L}(d | \theta_\text{ext})$, which describes the likelihood of observing gravitational-wave event data given the distribution of model parameters for an extremal event. 
In order to include measurement uncertainty, we define a new metric, ``the normalised evidence'':
\begin{align}\label{eq:normalised_evidence}
    \overline{\mathcal{Z}} \equiv & \overline{\pi(\theta_\text{ext} | M, \text{det}, N)}  \nonumber\\
    = & 
    \frac{ 
    \int d\theta \, {\cal L}(d | \theta_\text{ext}) \pi(\theta_\text{ext} | M, \text{det}, N)
    }{
    \int d\theta \,  {\cal L}(d | \theta_\text{ext}) \pi(\theta_\text{ext} | U)
    } , 
\end{align}
where the overline denotes an average over measurement uncertainty, and $\pi(\theta | U)$ is a prior uniform in the parameter of interest.\footnote{Astute readers may notice that $\overline{\mathcal{Z}}$ is equivalent to a Bayes factor comparing the model with the prior given in Eq.~\ref{eq:most_massive_detected} to a prior uniform in total mass.}
In the high signal-to-noise ratio (SNR) limit, the likelihood function becomes a delta function peaked at the true parameter values $\theta_\text{true}$, so
\begin{align}
    \overline{\mathcal{Z}} \rightarrow & 
    \frac{ 
    \int d\theta \, \delta(\theta_\text{ext} - \theta_\text{true}) \pi(\theta_\text{ext} | M, \text{det}, N)
    }{
    \int d\theta \,  \delta(\theta_\text{ext} - \theta_\text{true}) \pi(\theta_\text{ext} | U)
    } , 
\end{align}
which simplifies to
\begin{align}
    \overline{\mathcal{Z}} \rightarrow & 
    \frac{ 
    \pi(\theta_\text{true} | M, \text{det}, N)
    }{
    \pi(\theta_\text{true} | U)
    } .
\end{align}
In the high SNR limit, $\overline{\mathcal{Z}}$ approaches the ratio of prior probability densities at $\theta_\text{true}$.
Thus, the denominator of Eq.~\ref{eq:normalised_evidence} ensures $\overline{\mathcal{Z}}$ tracks the parameter of interest.
By constructing an empirical distribution of $\overline{\mathcal{Z}}$ with simulated data---denoted $\pi(\overline{\mathcal{Z}} | M, \text{det}, N)$---we can again calculate a $p$-value quantifying if the observed value ${\overline{\mathcal{Z}}}$ is unusual compared to the distribution expected from simulation:
\begin{align}\label{eq:Z_p_value}
    p \equiv & \int_{\Gamma_2} d\overline{\mathcal{Z}}' \, 
    \pi(\overline{\mathcal{Z}}'|M, \text{det}, N) \\
    \Gamma_2 = & \bigg\{\pi(\overline{\mathcal{Z}}'| M, \text{det}, N) < \pi(\overline{\mathcal{Z}} | M, \text{det}, N)\bigg\} .
\end{align}
If the observed value of total mass is unusual---after taking into account measurement uncertainty---then ${\overline{\mathcal{Z}}}$ will be small, and so the resulting $p$-value will be accordingly small, whereas if ${\overline{\mathcal{Z}}}$ is large, $p$ is ${\cal O}(1)$.
Moreover, this procedure reproduces Eq.~\ref{eq:naive-p} in the high signal-to-noise ratio  limit. 
For example, take a single observed event, at an SNR of either 10 or 1000. These events will have different support for the same model, described by their posterior distributions ``leaking" into the model distribution, and should therefore falsify this model to different degrees. It is this support that is captured by Eq. \ref{eq:Z_p_value}, and not by the maximum likelihood estimator in Eq.~\ref{eq:naive-p}.

\section{Demonstration} \label{results}
To demonstrate the above formalism, we test the ability of different models to explain the total mass $m_\text{tot}$ of the event GW190521.
We consider four models:
\begin{itemize}
    \item \citet{gayathriGravitationalWaveSource2023b} propose a model for binary black hole formation in an AGN disk with hierarchical mergers. 
    They adopt a one-parameter model parameterised by the maximum allowed natal black hole mass, $m_\text{max}$, while fiducial values of AGN parameters are chosen.
    A seed black hole distribution is generated following a Saltpeter mass function for each model (with a different value of $m_\text{max}$), and neutron stars are assumed to be normally distributed in mass with a mean of $1.49 M_\odot$ and a standard deviation of $0.19 M_\odot$.
    These compact objects are then allowed to migrate through an accretion disk around a supermassive black hole, encountering other objects and undergoing subsequent mergers.
    We consider three variations of this model, with $m_\text{max} = 15M_\odot, 50M_\odot \text{ and } 75 \text{M}_\odot$.
    These three models are labeled ``AGN.''
    \item \citet{rodriguezBlackHolesNext2019}'s model for binary black hole assembly in globular clusters. 
    They use a coupled modeling technique, in which stars are evolved from their zero-age main sequence births using a binary stellar evolution package \citep[][]{hurley_comprehensive_2000, hurley_evolution_2002, chatterjee_monte_2010}, while their movement through a globular cluster is also simulated using the \textsc{Cluster Monte Carlo} package \citep[][]{joshi_monte_2000, pattabiraman_parallel_2013}. Eventually, some of these stars become potential GW sources as they form compact objects and undergo binary formation and subsequent mergers. 
    In this Paper, we consider the model variation with natal black hole spin $\chi_{\text{birth}}=0$.
    This model is labeled ``Globular Cluster''.\footnote{While we abbreviate our two models as ``AGN'' and ``Globular Cluster,''
    it should go without saying that they many are other versions
    of binary formation models involving AGNs and globular
    clusters.}
\end{itemize}

For each model, we simulate $N=100$ events for the LIGO H1 and L1 observatories operating at design sensitivity.
We create $n=100$ catalogues of simulated events ($10^4$ events in total). 
We assign a spin magnitude to each component drawn from a Gaussian distribution centered at 0.7 and truncated at 0 and 1, to reflect their dynamical origin \citep[e.g.,][]{kimballBlackHoleGenealogy2020}.

Each event is injected into Gaussian noise colored by the amplitude spectral density noise curves of design-sensitivity LIGO. 
Each detection has network SNR $>12$.

To find the apparently most-massive event in each catalogue, we perform fast parameter estimation on the ten events with the largest true mass.
We use the nested sampler \textsc{dynesty} \citep[]{speagleDynestyDynamicNested2020a} as implemented in \textsc{Bilby} \citep[]{ashtonBilbyUserfriendlyBayesian2019} with only 100 live points, which is adequate to estimate the maximum likelihood value of total mass.
Our runs use the \textsc{IMRPhenomPv2} waveform approximant \citep[][]{schmidtModelsGravitationalWaveforms2012}. 
We choose boundary values of the component mass priors to include the range of masses in the injection set.
Due to noise fluctuations, the event with the largest true mass may not be the \textit{apparently} most-massive event, but tests suggest we can reliably find the apparently most-massive event among these ten.
We find that the truly most-massive event is the apparently most-massive event around $80\%$ of the time, while the second truly most-massive event is the apparently most-massive event around $10\%$ of the time. Few events below the third truly most-massive event are chosen as the apparently most-massive event. 

We use importance sampling to reweight the result \citep[see, e.g.,][]{intro} to obtain the marginal likelihood that we would have obtained using $\pi( \theta_\text{ext} | M_{\text{true}}, \text{det}, N)$---the distribution of truly most-massive events in each catalogue. 
We estimate $\pi( \theta_\text{ext} | M_{\text{true}}, \text{det}, N)$ by fitting the distribution of total mass and mass ratio posteriors of the truly most-massive events for each model with a kernel density estimator. 
We then calculate ${\overline{\cal Z}}$ (Eq.~\ref{eq:normalised_evidence}) for each event.
The value of ${\overline{\cal Z}}$ characterises how extreme each is compared to the distribution of truly most-massive events. 

However, a small value of ${\overline{\cal Z}}$ may indicate either an unusually low-mass event or unusually high-mass event.
Therefore, we select the apparently most-massive event in each catalogue as follows: if any events have a higher max-likelihood total-mass estimate greater than the median total mass of $\pi( \theta_\text{ext} | M_\text{true}, \text{det}, N)$, we select the event of these with the lowest ${\overline{\cal Z}}$ to be the apparently most massive. 
If no such events exist, we select the event with the highest max-likelihood total-mass estimate to be the apparently most massive.
This ensures that we do not select low-mass outliers.
We demonstrate this method in further detail in Appendix \ref{appendixa}.

\begin{figure*}
    \centering
    \includegraphics[width=180mm]{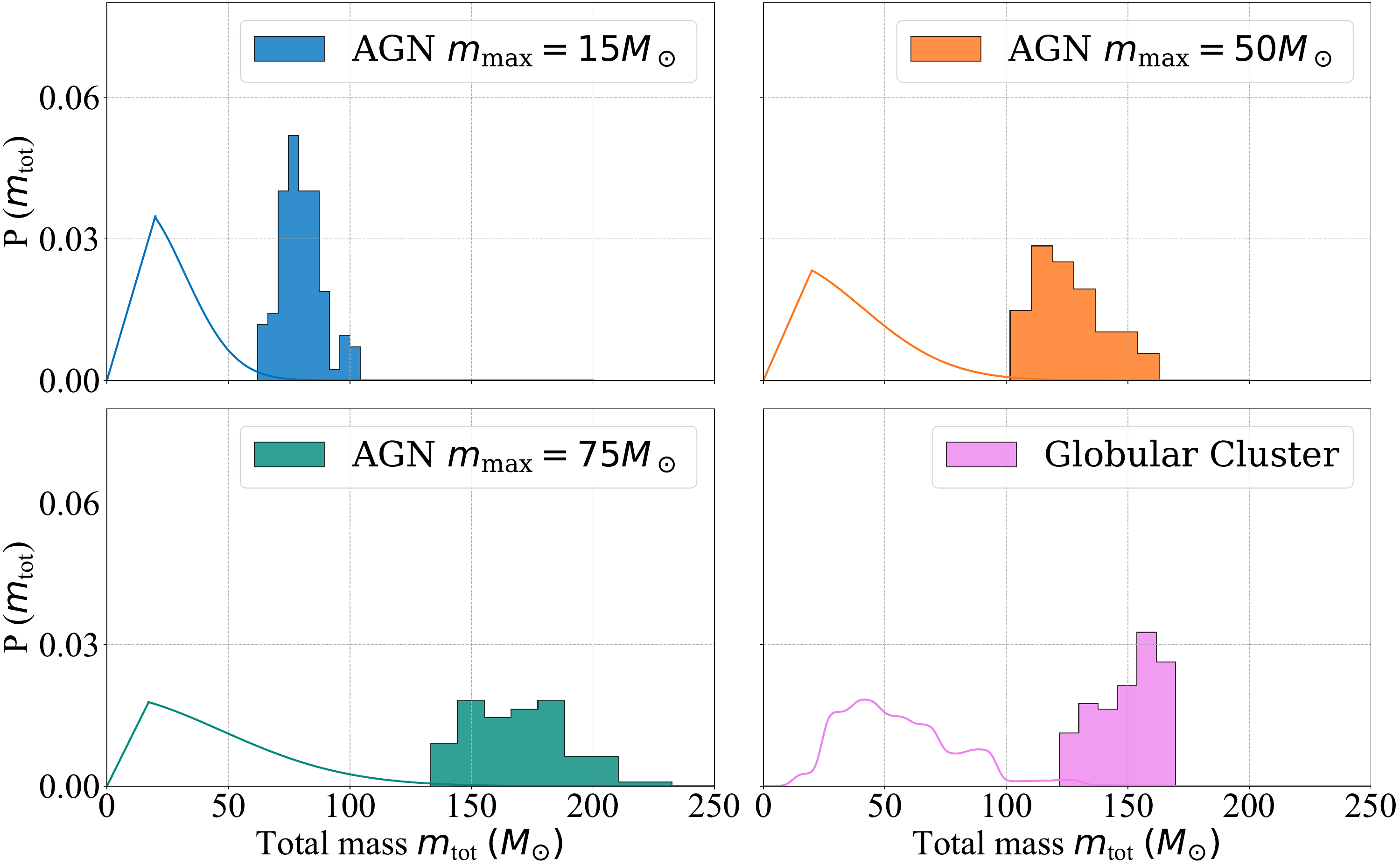}
    \caption{
    The distribution of total mass $m_\text{tot}$ for different models.
    The solid curves are the original models while the histograms show the {total-mass distribution of the  of the apparently most-massive event in each simulated catalogue.}
    We consider three variations of the AGN model from \citet{gayathriGravitationalWaveSource2023b} with different maximum black hole masses $m_\text{max}$ (blue, orange, green).
    We include a globular cluster model from \citet{rodriguezBlackHolesNext2019} (pink).}
    \label{fig:combined_total_mass}
\end{figure*}

The distributions of maximum total mass for each model are shown in Fig.~\ref{fig:combined_total_mass}. For the apparently most-massive event in each catalogue, we calculate ${\overline{\cal Z}}$ (Eq.~\ref{eq:normalised_evidence}) with the nested sampler \textsc{dynesty} as implemented in \textsc{Bilby} with 1000 live points, using the same noise realisation for each event as before. 
We use importance sampling to reweight the result to obtain the marginal likelihood that we would have obtained using $\pi( \theta_\text{ext} | M_{\text{app}}, \text{det}, N)$, the distribution of apparently most-massive events, which we once again estimate using a kernel density estimator.

\begin{figure*}
    \centering
    \includegraphics[width=90mm]{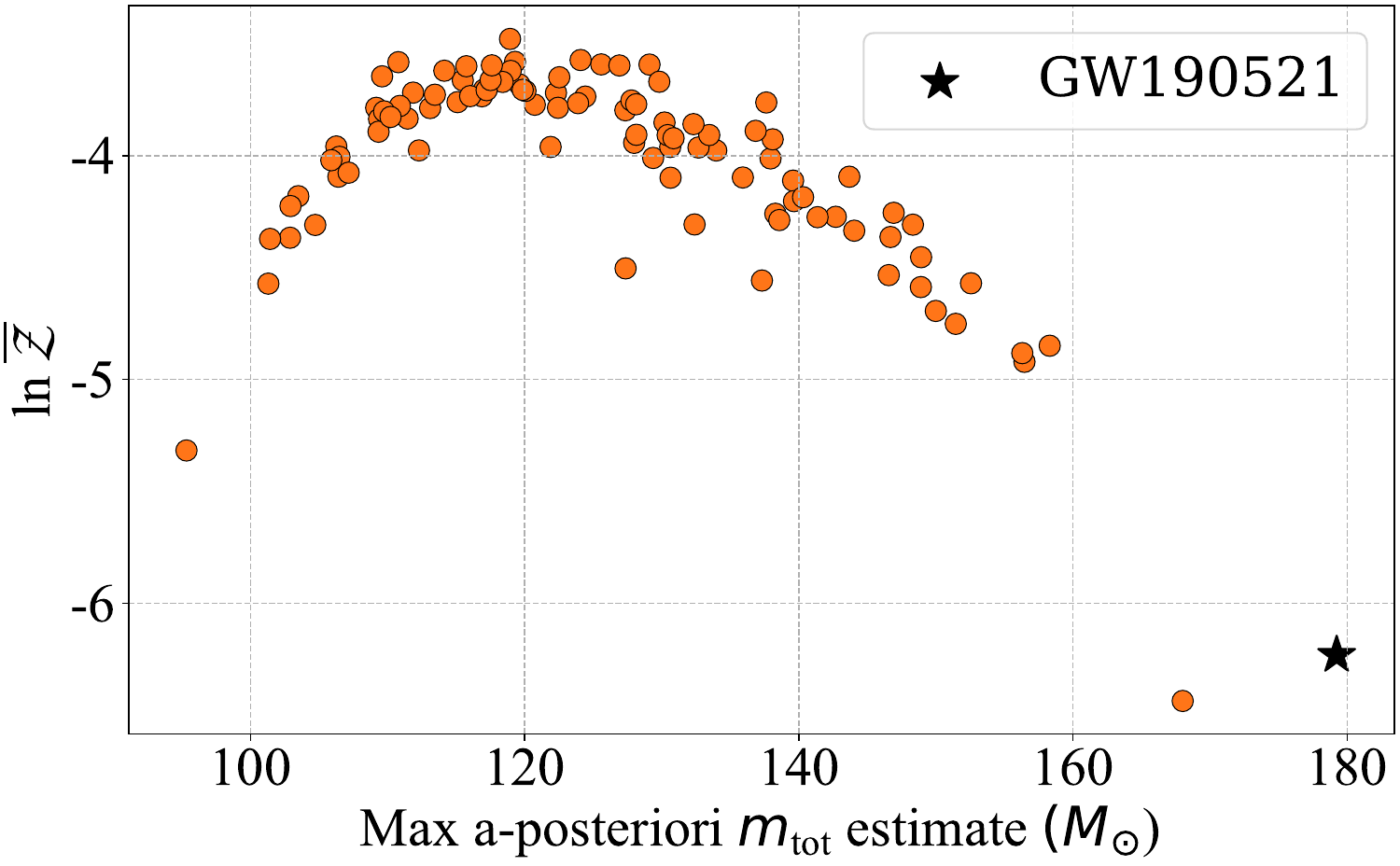}
    \caption{Log-normalised evidence $\ln{\overline{\mathcal{Z}}}$ versus max-a-posteriori total-mass estimate for the apparently most-massive events of the $m_\text{max} = 50 M_\odot$ AGN model. 
    Lower $\ln{\overline{\mathcal{Z}}}$ values correspond to catalogues that are relatively unusual---either because the maximum total mass is either unusually small or unusually large.
    The highest values of $\ln{\overline{\mathcal{Z}}}$ correspond to typical catalogues with usual values for the maximum total mass.
    The log-normalised evidence for GW190521, calculated using this model, is also plotted against the max-a-posteriori total mass of GW190521.}
    \label{fig:lnX_vs_total_mass}
\end{figure*}

As an example, in Fig.~\ref{fig:lnX_vs_total_mass} we plot $\ln\overline{\cal Z}$ versus the maximum-likelihood value of $m_\text{tot}$ for the apparently most-massive event in each simulated catalogue, for the $m_\text{max} = 50 M_\odot$ AGN model.
The most common value of $m_\text{tot}$ is $\approx\unit[130]{M_\odot}$, consistent with expectations from Fig.~\ref{fig:combined_total_mass}.
As expected, these typical values of $m_\text{tot}$ tend to produce relatively large values of $\ln\overline{\cal Z}$.

\begin{figure*}
    \centering
    \includegraphics[width=180mm]{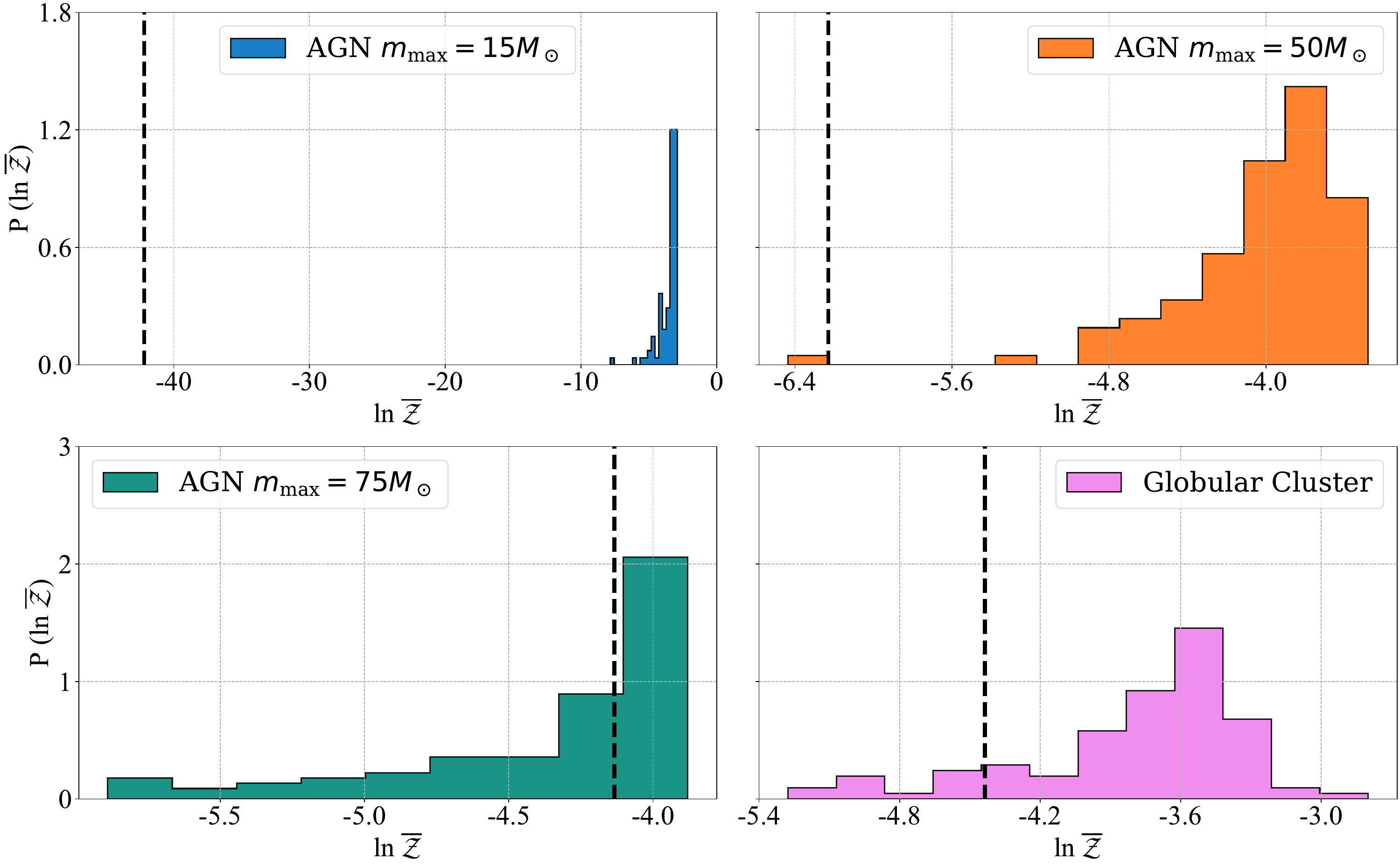}
    \caption{
    Probability distribution of log-normalised evidence $(\ln{\overline{\mathcal{Z}}})$ for distributions of the apparently most-massive, detectable events drawn from Eq.~\ref{eq:most_massive_detected} (histograms), against $\ln{{\overline{\mathcal{Z}}}'}$, the log-normalised evidence for GW190521 (black-dashed line), for (from left to right, top to bottom) AGN models with $m_\text{max} = 15 \text{} M_\odot  \text{ (blue), } 50 \text{} M_\odot \text{ (orange) }\text{ and } 75 \text{} M_\odot \text{ (green)}$, as well as a globular cluster model with $\chi_\text{birth} = 0$ (pink).
    The $p$-value calculation in Eq.~\ref{eq:Z_p_value} is the integral over the region to the left of $\ln{{\overline{\mathcal{Z}}}'}$.
    }
    \label{fig:combined_hist}
\end{figure*}

Continuing with the $m_\text{max} = 50 M_\odot$ AGN model as an example, we use the empirical distribution of $\ln\overline{\cal Z}$ (shown in orange in Fig.~\ref{fig:combined_hist}) to calculate a $p$-value for GW190521 ({$\ln\overline{\cal Z}=-6.23$}; black-dashed line).
Since 99\% of the simulations produce $\ln\overline{\cal Z}$ values larger than the one obtained for GW190521, the $p$-value is {$p=0.01$} (disfavored at the two-sigma level).
This implies the total mass of GW190521 is very unusual (at the two sigma level) compared to the distribution of expected apparently most-massive events after $N=100$ events and given the \cite{gayathriGravitationalWaveSource2023b} model.

We repeat this process for the $m_\text{max} = 15M_\odot \text{ and } 75 M_\odot$ AGN models, as well as the globular cluster model. 
We summarise the results in Table \ref{tab:p-value}; see also Fig.~\ref{fig:combined_hist}. 
We find {$p\simeq0$} for the $m_\text{max} = 15 M_\odot$ AGN model, indicating this model almost never produces an event as massive as GW190521. We find the $m_\text{max} = 75 M_\odot$ AGN model has {$p = 0.61$}, suggesting the model produces GW190521-like events often. 
Finally, we find the globular cluster model has {$p = 0.12$}, suggesting that it is reasonable to draw a GW190521-like event from this model.

\begin{table}
    \centering
    \caption{Associated $p$-value for each model based on AGN natal black hole mass $m_\text{max}$ or globular cluster (GC) scenario.}
    \label{tab:p-value}
    \begin{tabular}{lcccc}
        \hline
        Model & $15 M_\odot$ & $50 M_\odot$ & $75 M_\odot$ & GC \\
        \hline
        $p$-value & $\sim0$ & $0.01$ & $0.61$ & $0.12$ \\
        \hline
    \end{tabular}
\end{table}

\section{Discussion}\label{discussion}
We have demonstrated the use of our statistical framework, which makes use of the ``normalised evidence'' $\overline{\mathcal{Z}}$ in testing several models of binary formation using the exceptional event GW190521. 
We rule out the AGN model with $m_\text{max} = 15 \text{} M_\odot$.
The AGN model with $m_\text{max} = 50 \text{} M_\odot$ is disfavoured in explaining GW190521 (at the two-sigma level).
The AGN model with $m_\text{max} = 75 \text{} M_\odot$ and the globular cluster model plausibly provide adequate explanations for GW190521.

In order to compare our method with the $m_\text{tot}$ max-likelihood-estimated $p$-value calculation from \cite{fishbachMostMassiveBinary2020a}, we carry out a simulation. 
We consider outlier events for the $m_\text{max} = 50 M_\odot$ AGN model and consider two different situations.
In both cases, we observe an outlier in total mass with maximum-likelihood value of $m_\text{tot} = 165 M_\odot$.
However, in one case, the network signal-to-noise ratio is relatively low (SNR $=12$) while in the other case the signal-to-noise ratio is high (SNR $=45$).

In the \cite{fishbachMostMassiveBinary2020a} framework, these two events are assigned the same $p$-value because they both have the same maximum-likelihood value for $m_\text{tot}$.
These $p$-values, however, do not fold in all the available information.
Since the posterior for $m_\text{tot}$ narrows with increasing signal-to-noise ratio, we can be relatively more confident that the SNR $=45$ event is incompatible with the AGN model because the total mass is more assuredly measured outside the predicted range of the model.
Since our approach includes information about the likelihood width, the SNR $=45$ event yields a lower $p$-value than the SNR $=12$ event.

\begin{figure*}
    \centering
    \includegraphics[width=90mm]{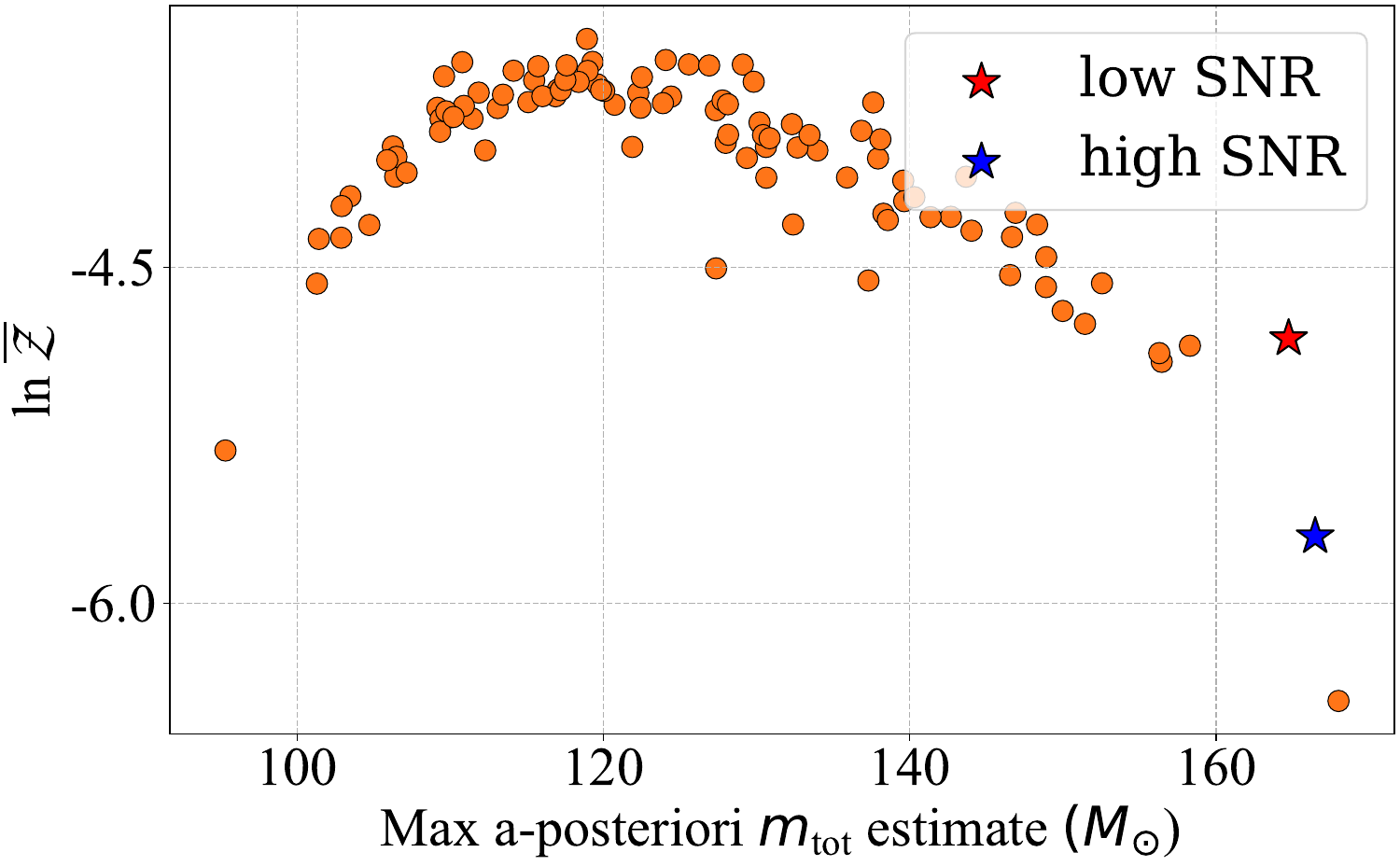}
    \caption{Log-normalised evidence $\ln{\overline{\mathcal{Z}}}$ versus max-a-posteriori total-mass estimate for the apparently most-massive events of the $m_\text{max} = 50 M_\odot$ AGN model, as in Fig. \ref{fig:lnX_vs_total_mass}.
    The red and blue stars indicate the $\ln{\overline{\mathcal{Z}}}$ calculated for an event with the same peak total-mass likelihood of $165 M_\odot$, but with a network SNR of $12$ and $45$. The low-SNR event's posterior leaks into the model distribution to a greater extent than the high-SNR event, and is therefore assigned a higher $p$-value. The high-SNR event is more confidently in the tail of the distribution, with a lower associated $p$-value.}
    \label{fig:lnX_vs_total_mass_low_high}
\end{figure*}

To illustrate, we inject two events with different distances into coloured Gaussian noise and perform parameter estimation as above. The results are illustrated in Fig. \ref{fig:lnX_vs_total_mass_low_high}. 
The SNR $=45$ event yields $p = 0.01$, indicating the $m_\text{max} = 50 M_\odot$ AGN model poorly explains the event.
The SNR $=12$ event yields $p=0.05$, which also disfavours the $50 M_\odot$ AGN model, but with less statistical significance.

For future work we propose to apply this method more broadly in order to falsify models as explanations for exceptional events.
Key questions include:
\begin{itemize}
    \item Which models are consistent with the most massive binary black holes?
    \item Which models are consistent with the most extreme spins?
    \item Which models are consistent with the most extreme mass ratios, such as GW190814~\citep{theligoscientificcollaborationGW190814GravitationalWaves2020}?
    \item Similarly, which models are consistent with the smallest secondary masses, like the $m_2=2.6 M_\odot$ object observed in GW190814~\citep{theligoscientificcollaborationGW190814GravitationalWaves2020}?
\end{itemize}
There are many ways to evaluate the explanatory power of different population models, each of which provides the answer to a different question.
The Bayes factor can be used to determine which of two population models provides a better description of the data \citep[e.g.][]{mouldOneManyComparing2023}. We note it may be useful to determine if none of the models tested provide an adequate fit---something we do not learn from the Bayes factor, which is a relative comparison of models.

Leave-one-out analyses can be used to determine if one event is a statistical outlier with respect to the rest of the population \citep[e.g.][]{essickProbingExtremalGravitationalWave2022}.
Our method may be similarly extended to examine whether fits to the entire observed population can explain the most exceptional events observed. If the model is fit to the data and still struggles to explain an exceptional event, it is clearly inadequate.

Finally, the ``maximum population likelihood'' may be used to test a model's goodness of fit to observed events \citep[]{Lstroke}. This method characterises a model's fit to an entire catalogue of events, requiring parameter estimation on all events in many simulated catalogues. 
Our method focuses only on exceptional events, and thus vastly reduces the total computation required.

Our framework is an example of ``model criticism'' \citep[][]{romero-shawWhenModelsFail2022}. It is a posterior predictive check in which one infers a model distribution from observed data, then compares samples from that distribution back to data, to check for consistency \citep[e.g.,][]{fishbachDoesMatterMatter2020a}. Specifically, we perform a posterior predictive check on a model for a single observed extreme event. This requires adopting a frequentist approach, as opposed to an entirely Bayesian framework, as well as comparisons of events' ``normalised evidences'', rather than comparisons of their posteriors.

We note that a similar approach to model criticism was recently adopted by \citet[]{amendolaDistributionBayesRatio2024}, who developed a model-testing framework using a $p-$value calculated over a distribution of Bayes factors. They applied this statistic to cosmological supernovae 1a data, and found their method similarly avoids shortcomings of purely Bayesian or frequentist approaches. The multi-disciplinary applicability of such a statistic highlights its potential to improve model testing not only in gravitational wave data, but across diverse scientific fields.

\section*{Acknowledgements}
We thank Thomas Callister, Matthew Mould and the referee for their helpful comments.
This is LIGO document \#P2400175.
We acknowledge support from the Australian Research
Council (ARC) Centres of Excellence CE170100004 and CE230100016, as well as ARC
LE210100002, and ARC DP230103088. L.P.
receives support from the Australian Government Research
Training Program. 
S.S is a recipient of an ARC Discovery Early Career Research Award (DE220100241).
This material is
based upon work supported by NSF’s LIGO Laboratory
which is a major facility fully funded by the National
Science Foundation. The authors are grateful for computational resources provided by the LIGO Laboratory
and supported by National Science Foundation Grants
PHY-0757058 and PHY-0823459.

This research has made use of data or software obtained from the Gravitational Wave Open Science Center (gw-openscience.org), a service of LIGO Laboratory,
the LIGO Scientific Collaboration, the Virgo Collaboration, and KAGRA. LIGO Laboratory and Advanced
LIGO are funded by the United States National Science Foundation (NSF) as well as the Science and Technology Facilities Council (STFC) of the United Kingdom, the Max-Planck-Society (MPS), and the State of
Niedersachsen/Germany for support of the construction
of Advanced LIGO and construction and operation of
the GEO600 detector. Additional support for Advanced
LIGO was provided by the Australian Research Council.
Virgo is funded, through the European Gravitational
Observatory (EGO), by the French Centre National
de Recherche Scientifique (CNRS), the Italian Istituto
Nazionale di Fisica Nucleare (INFN) and the Dutch
Nikhef, with contributions by institutions from Belgium,
Germany, Greece, Hungary, Ireland, Japan, Monaco,
Poland, Portugal, Spain. The construction and operation of KAGRA are funded by Ministry of Education,
Culture, Sports, Science and Technology (MEXT), and
Japan Society for the Promotion of Science (JSPS), National Research Foundation (NRF) and Ministry of Science and ICT (MSIT) in Korea, Academia Sinica (AS)
and the Ministry of Science and Technology (MoST) in
Taiwan.
\section*{Data Availability}

The data underyling this article are publicly available at \url{https://www.gw-openscience.org}.



\bibliographystyle{aasjournal}
\bibliography{refs} 



{
\appendix
\section{Choosing the apparently most-massive event in each catalogue}\label{appendixa}
Choosing the apparently most-massive event in each catalogue requires a ranking scheme. 
In this paper, we identify outliers using the quantity ${\overline{\mathcal{Z}}}$.
Thus, this metric can be used to identify the apparently most-massive event in each catalogue. 
However, small values of ${\overline{\mathcal{Z}}}$ can indicate two kinds of outliers: events with an unusually high mass and events with an unusually low mass. 
Thus, we must take care when identifying the apparently most-massive event in each catalogue. 
We adopt the following ranking scheme:
\begin{enumerate}
    \item If there is one or more event with a total mass greater than the median total mass of the distribution of truly most-massive events, we select among this subset of events the event with the lowest ${\overline{\mathcal{Z}}}$ to be the apparently most-massive.
    \item If there are no such events (with a total mass greater than the median total mass of the distribution of truly most-massive events), we select the event with the highest max-likelihood total-mass estimate to be the apparently most massive.
\end{enumerate}

We demonstrate this scheme in Fig.~\ref{fig:three_plots_apparent}, using events drawn from the AGN model with $m_\text{max} = 50M_\odot$. 
In the left panel, we show the case where none of the top ten truly most-massive events in a catalogue have a max-likelihood total-mass estimate above the median of the total-mass distribution for the truly most-massive events in each catalogue.
(By construction, this is the case for 50\% of our catalogues.)
In this case, we take the event (denoted with peach shading) with the highest max-likelihood total-mass estimate to be the apparently most-massive. 

In the middle panel, we show the case where several events exist above the aforementioned median (indicated with a dashed line). In this case, we take the (peach-coloured) event to the right of the dashed line with the lowest $\ln{\overline{\mathcal{Z}}}$ to be the apparently most-massive. 
In this case, the event with the highest maximum-likelihood mass is also the apparently most-massive event.
In the right panel, we again take the (peach-coloured) event to the right of the median with the lowest $\ln{\overline{\mathcal{Z}}}$ to be the apparently most-massive.
In this catalogue, the second most-massive event by max-likelihood estimate is the apparently most-massive event. 

\begin{figure*}[h]
    \centering
    \includegraphics[width = 180mm]{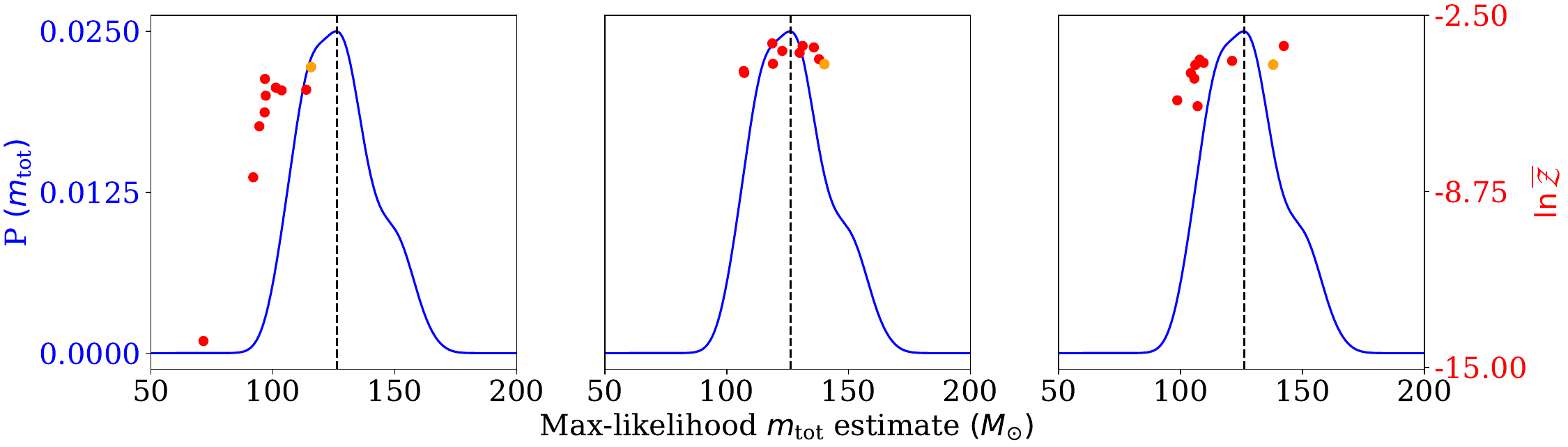}
    \caption{Dual plot showing the calculated $\ln{\overline{\mathcal{Z}}}$ for the top 10 truly most-massive events from three catalogues of the AGN model with $m_\text{max} = 50M_\odot$ (red), with the apparently most-massive event for each catalogue shown in peach. The quantity $\ln{\overline{\mathcal{Z}}}$ is calculated using the distribution of truly most-massive events,  $\pi( \theta_\text{ext} | M_{\text{true}}, \text{det}, N)$ (blue). The median total mass of this distribution is shown as the black dotted line. The left panel shows the case where none of the top 10 truly most-massive events have a max-likelihood total-mass estimate above the median total mass of $\pi( \theta_\text{ext} | M_{\text{true}}, \text{det}, N)$, and thus the apparently most-massive event is the event with the highest max-likelihood total-mass estimate. The middle panel shows the case where several events have a higher max-likelihood total-mass estimate than the aforementioned median, and thus the event with the lowest $\ln{\overline{\mathcal{Z}}}$ is the apparently most-massive. In the right panel, two events have a higher max-likelihood total-mass estimate than the aforementioned median, and thus the event with the lowest $\ln{\overline{\mathcal{Z}}}$ is the apparently most-massive. Importantly, this event does not have the highest max-likelihood total-mass estimate.}
    \label{fig:three_plots_apparent}
\end{figure*}
}


\end{document}